\documentstyle[preprint,aps]{revtex}

\begin{document}
\input epsf

\title{Macroscopic quantum tunneling and resonances in coupled
Bose-Einstein condensates with oscillating atomic scattering length}
\author{F.Kh.\ Abdullaev\thanks{On leave of absence from the 
Physical-Technical Institute, Tashkent, Uzbekistan}, R.A. Kraenkel} 
\address{Instituto de F\'{i}sica T\'eorica, Universidade Estadual
Paulista,\\
Rua Pumplona 145,01405-900, S\~ao Paulo, Brasil\\}
\date{\today}
\maketitle

\begin{abstract}
We study the macroscopic quantum tunneling , self-trapping
phenomena in two weakly coupled Bose-Einstein condensates with
periodically time-varying atomic scattering length.

The resonances in the oscillations of the atomic populations are
investigated. We consider oscillations in the cases of macroscopic quantum
tunneling and
the self-trapping regimes.  The existence of chaotic
oscillations in the relative atomic population due to overlaps between
nonlinear resonances is showed. We derive the
whisker-type map for the
problem and obtain the estimate for the critical amplitude of
modulations leading to chaos. The diffusion coefficient for motion in
the stochastic layer near separatrix is calculated. The
analysis of the oscillations in the rapidly varying case shows
the possibilty of stabilization of the unstable $\pi$ - mode regime.

\end{abstract} 

\newpage
\section{Introduction}

Recently  much attention has been paid to the investigation
of macroscopic tunneling phenomena between coupled
Bose-Einstein condensates (BEC) \cite{Andrews,Dalf}. The theory has been 
developed  for weakly as well as for strongly overlapped
condensates \cite{Coley}-\cite{Ostrovskaya}. As a result, it was shown that, 
in oscillations of the relative atomic population,  periodic processes of two
types can exist. One which is characterized by a zero mean  
atomic inbalance function 
$(<z(t)> = 0)$, corresponding to the periodic flux of atoms between two
condensates due to the overlaps of the wavefunctions. 
The second type has a net atomic population balance $<z(t)> \neq 0$, 
corresponding to the so-called macroscopic self-trapping , with localization 
of atoms in one of the
condensates, and a periodically varying population around a constant 
value.
The different regimes are connected with the value of the initial phase
differences $\theta = \phi_1 - \phi_2$ between the condensates and the
effective nonlinearity $\Lambda \sim (\alpha_1 + \alpha_2 )N$, where
$\alpha_i$
are the nonlinearity parameters depending on the atomic scattering length $a_s$ 
(which in turn depends on the two-atoms interactions), and $N$ is the total number 
of atoms.

If a periodic variation of the parameters of the trap and of the BEC is 
introduced, new phenomena show up. Examples of this are 
the resonant phenomena with a 
periodically varying trap potential considered recently for the case of 
weakly coupled BEC in
\cite{Kraenkel} and  for the strongly overlaped BEC in \cite{Elyutin}. 
The optical analog of this process is the electromagnetic wave
transmission in a nonlinear coupler with periodic variation of tunnel
coupling \cite{Darmanyan,Malomed}.
Further interesting cases occur when a time-dependent atomic scattering
length in BEC is considered. This is made possible due to the fact that $a_s$
depends resonantly on the external magnetic field or on the laser field
\cite{Fedichev}-\cite{Cornish}.
This means that the scattering length can be varied in time, in particular,
periodically. The main interest now is to investigate the influence of this
periodic variation of atomic scattering length on the macroscopic
interference phenomena. The central point here consists in that, as the 
interference phenomena are
time-periodic processes in a nonlinear system, we
can expect a nonlinear resonant
response of the system under  time-periodic perturbations, leading to a 
variety of phenomena, such as hysteresis or chaos.

\section{Formulation of problem. Slowly time-varying atomic
scattering length}

The problem of weakly coupled BEC's in a double-well  trap potential
with time-dependent atomic scattering length $a_s$ can be
described by the two-modes model
\begin{eqnarray}\label{1}
ih\frac{\partial \psi_1}{\partial t} = (E_1 + \alpha_1 (t) |\psi_1 |^2
)\psi_1
 - K\psi_2 ,\nonumber \\
ih\frac{\partial \psi_2}{\partial t} = (E_2 + \alpha_2 (t)\psi_2 |^2
)\psi_2 - K\psi_1 ,
\end{eqnarray}
where the parameters  
$E_i, \alpha_i , K$ are defined by the overlaps integrals of the time
dependent Gross-Pitaevsky eigenfunctions and derived in  \cite{Shenoy1}. In 
particular,  
$\alpha_i (t)$ are the  parameters characterizing the 
nonlinear interactions between atoms and $\alpha_i \sim g_0 = 4\pi h^2
a_{s}(t)/m$, where
$a_{s}(t)$ is the time dependent atomic scattering length. 
This system of equations is valid in
the approximation of a weak link between condensates. As  
comparison with the numerical solution of GPE shows that it is a good 
aproximation  for $z \leq 0.6.$ (see the definition of $z(t)$ below).  Anther 
limitation is connected 
with the time dependence of $a_s$. For a harmonic modulation
in time we should require that the perturbation does not  introduce 
transitions between the ground state and excited states in the
trap, whose energy gap  is of  the order of $\hbar \omega_0$, where $\omega_0$ 
is the harmonic frequency of trap. Then we have the restriction $\Omega \ll 
\omega_0$. 
Another characteristic frequency is given by the coupling between
condensates and is defined as $\omega_L = 2K$. For the weak coupling case
when  $\omega_L \ll \omega_0$ the regimes of resonant $\Omega \sim
\omega_L$ and rapid modulations of $a_s$ can be realized.
The periodic modulation of nonlinearity in the Gross-Pitaevsky equation can 
also lead to the parametric instability of the collective excitations
of the condensate \cite{Matera,Abdullaev2}, that, in turn,  leads to the break 
down of the two-modes
approximation. Because these frequencies are far from the ones considered below 
we wiil use the model (\ref{1}) in our analysis.

Introducing new variables $\psi_i = \sqrt{N_i}\exp(i\theta_i ),  z = (N_1
-
N_2)/N_T , N_T =
N_1 + N_2 , \Phi = \theta_1 -
\theta_2$, where $N_i, \theta_i$ are the number of atoms and phases in the i-th
trap, we get the system
\begin{eqnarray}\label{tun}
z_t = -\sqrt{1-z^2}\sin{\Phi},\\
\Phi_t =\nu \Lambda(t) z + \frac{z}{\sqrt{1-z^2}}\cos{\Phi} + \Delta
E_0,
\end{eqnarray}
where $\Lambda = (\alpha_{1}(t)+ \alpha_{2}(t) ) N_T/2K$, $t = \omega_{L} t$,
$\omega_L = 2K$, $\omega_L$ being the frequency of the linear Rabi
oscillations, $\Delta E_0 = \Delta E/2K$,  and $ \nu = \pm
1 $
for the 
positive and negative atomic scattering length, respectively. 

The Hamiltonian of the unperturbed system ($a_{s}(t) = {\rm const}$, $\Delta E_0 
= {\rm const}$, $\eta = 0$) is
\begin{equation}
H = \frac{\nu \Lambda z^2}{2} - \sqrt{1-z^2}\cos(\Phi) + \Delta E_0 z.
\end{equation}
Below, if not specified otherwise, we will consider the case of positive 
scattering
length $\nu = 1$,and $ \Delta E_0 = 0$.
Consider the initial phase difference to be zero, $\Phi(0) = 0$ -- the zero-phase 
mode. Depending 
on the number of atoms, $N$, and two-atoms interaction, $\alpha$, i.e.
$\Lambda$, there may exist a macroscopic quantum tunneling regime with $<z(t)> = 
0$,corresponding to $\Lambda < \Lambda_c$ and a self-trapped regime with  $<z(t)> 
\neq
0$, corresponding to $\Lambda > \Lambda_c$. The same is valid for the initial 
phase
difference,  $\Phi(0) = \pi$ -- the $\pi$-phase mode. 

Let us consider now the case of periodic modulations of the atomic scattering
length, when $a_{s}(t) = a_0(1 + \epsilon 
\cos{\Omega t})$, inducing a variation of $\Lambda$, i.e,  $ \Lambda =
\Lambda_0 
+ 
\Lambda_1\cos(\Omega t)$. Such a variation can be achieved by the
variation of the external magnetic field \cite{Cornish} - a magnetic field 
induced Feshbach resonances for example. Near the resonance the scattering
length is varied dispersively and can have positive and negative values. 
We will consider situations when we are not close to resonance and
variations of scattering length are small, that is, $\Lambda_1/\Lambda_0 = 
\varepsilon \ll 1$. 


\subsection{Resonances in zero phase modes}

Let us first consider the case where $\Phi(0) = 0$. Take the
case of 
$z^2 \ll 1$ and $\Lambda_1 > z$. Taking
into account that $z_t \sim \Phi \sim z^2$, assumed to be small,
we obtain the equation for the relative atomic population
\begin{equation}
z_{tt} + (\Lambda(t) + 1)z - \frac{\Lambda(t) + 1}{2}z^3 = 0.
\end{equation}
where, $\Lambda(t) = \Lambda_0 (1 + \epsilon \sin(\Omega t))$.
For $\epsilon \ll 1$ we can reduce the equation to 
\begin{equation}
z_{tt} + \omega_0^2 z  - \beta z^3 + \epsilon \Lambda_0 
\cos(\Omega t)z
= 0,
\end{equation}
where $\omega_0 = \sqrt{1 + \Lambda_0}$ is the frequency of linear
oscillations, and $\beta = (\Lambda_0 + 1)/2$.
It is useful to introduce a new variable $y = \sqrt{(\Lambda_0 + 1)}z \le 1$
and a scaled 
time $\tau = t\omega_0$. Then we have the equation 
\begin{equation}
y_{\tau\tau} + y  - \beta_1 y^3 + \varepsilon_1
\cos(\Omega_1 \tau )y = 0,
\end{equation}
where $ \beta_1 = 1/(2\sqrt{\Lambda_0 
+ 1}), \varepsilon_1 = \varepsilon \Lambda_0/(\Lambda_0 + 1), \Omega_1 =
\Omega/\omega_0$.
  
Let us consider the dynamics   in the parametric resonance region when 
$\Omega_1 = 2 + \Delta$. Using the
results of perturbation theory
\cite{Landau}, we find that the parametric resonance occurs when the
condition for the square of increment $s^2 > 0$ is satisfied, where
\begin{equation}
s^2 = \frac{1}{4}[(\frac{\varepsilon_1}{2}(1 - \frac{3\beta_1 a^2}{8}))^2 
- \Delta^2 ].
\end{equation}
We have included here a nonlinear correction to the frequency $3\beta_1
a^2/8$. Then we obtain the limit for the amplitude
for parametric
resonance to occur,
\begin{equation}
a_p^2 \le \frac{8(\varepsilon_1 - 2\Delta)}{3\beta_1 \varepsilon_1} \quad , 
\end{equation}
and $z_p = a_p /\sqrt{\Lambda_0 + 1}$.

In Fig.(1) we present the oscillations of $z(t)$ when the parameters are in
the region of the parametric resonance, by integrating numerically Eq.(\ref{tun}).


\subsection{Resonances in $\pi$ - phase modes}

We now come to the case $\phi(0)=\pi$, with  for $\Lambda < 1$ and $z^2 <<
1$. 
The system (\ref{tun}) is simplified and reduces to one equation for
$\Phi$, 
having
the form of the equation of a double pendulum. Inspection of this equation 
shows that there exists a valley in the effective energy $V(\Phi)$ around $\Phi 
= \pi$. The maximum of the valley depth is achieved when $\Lambda_0 \rightarrow 
1$. Thus we can search the solution
of the system in this region of parameters assuming $z \ll 1, \Phi = \pi +
\delta(t)$. It results that we obtain the following equation for $z(t)$
\begin{equation}
z_{tt} + (1 - \Lambda_0)z - \Lambda_1 \cos(\Omega t)z = 0.
\end{equation}
Here $\Lambda_0 < 1$ .The parametric resonance in the oscillations
of the atomic population appears when $\Omega = 2\sqrt{1 - \Lambda_0} + \Delta$.
The
parametric instability occurs when $ \Lambda_1/(2\sqrt{1 - \Lambda_0})
\ge \Delta$. The increment  $s$ is, in this case,  
\begin{equation}
s = \frac{1}{2}\sqrt{(\frac{\Lambda_1}{2\sqrt{1- \Lambda_0}})^2  -
\Delta^2}.
\end{equation} 

For example, when $\Lambda_0 = 0.36, \Delta = 0.1$ we have $\Omega = 1.7$.
In Fig.(2) we plot the oscillations of $z(t) $ for this choice of
parameters, and with $\Lambda_1 = 0.162$, by a nummerical integration of Eq.(\ref{tun}). 

Note that the frequency of linear Rabi oscillations is $\omega_L = 2K$, or one in 
the units used. Thus the
resonance in
$\pi$ -mode occurs at frequencies {\it lower} than the Rabi frequency.
It is worth mentioning  that the metastable $\pi$-phase mode has been recently 
observed in the weak link separating two reservoirs of superfluid He$^3$
\cite{Backhaus}.

\section{Chaotic dynamics in oscillations of the relative atomic
population}

The oscillations induced by the periodic perturbations of the scattering length 
may be comples and become chaotic, for certain regions of the parameters. Note 
that the unperturbed system is equivalent to the Duffing oscillator.
Thus, as is known,  the resonance overlaps under periodic perturbations can lead 
to chaotic variations of $z(t)$ \cite{Holmes,Reichl,Abdullaev}.
In such cases, to study the chaotic motion, it is useful to look for the motion 
near  the separatrix of the unperturbed system. 
The total Hamiltonian is $H = H_0 + V$, where $H_0$ is unperturbed
Hamiltonian (4) and the perturbation is $V = \Lambda_1 \cos(\Omega
t)z^2/2 $. The value of the 
Hamiltonian on  separatrix is $H_s = 1$. The separatrix divides
the regions with macroscopic quantum tunnelling, with $<z> = 0$, from the 
self-trapped regions where 
$<z> \neq 0$. The expressions for the separatrix solution are: 
\begin{eqnarray}\label{sep}
z_s(t) = \sqrt{\frac{a}{b}}\mbox{sech}(\sqrt{2a}t),\nonumber \\
 \sin^2 (\Phi_s) = \frac{a^2 \mbox{sech}^2 (\sqrt{2a} t)\tanh^2
(\sqrt{2a}t)}{2K_0^2 (b
- a \mbox{sech}^2 (\sqrt{2a}t))},  
\end{eqnarray}
where $a = |1 - \Lambda |/2$, $b = \Lambda^2 /8$.
For example, for the zero-phase mode the separatrix is given by the hyperbolic 
fixed
points $\Phi = \pm\pi, z = 0$.

The existence of chaos can be proved by  calculating  the Melnikov
function. The calculations are similar to the ones performed in \cite{Kraenkel} 
and
shows that the Melnikov function has an the infinite
number of zeros, so  the existence of chaotic regimes in the atomic
population
oscillations is proved. 
In Fig.(3) we plot the Melnikov distance versus the frequency $\Omega$ for
$\Lambda = 10$. We see that the maximum value is achieved for $\Omega
\approx 3.89$.
In Fig.4 we plot the difference of atomic population $z(t)$ as function
of time for $\Lambda_0 = 10, z(0) = 0.4, \Lambda_1 = 0.2, \Omega = 3.89$. 

It is interesting to obtain an analytical estimate for the critical
amplitude of modulation $\Lambda_1$ leading to the chaotic behavior in the
relative atomic population. Using the expression for the separatrix
(\ref{sep}) we can calculate the 
Melnikov-Arnold integral and  find that the energy change for a
perturbation $\Lambda_1 \cos(\Omega t + \psi)$(where $\psi$ is the phase)
is
\begin{equation}\label{MA}
\Delta H = \int_{-\infty}^{\infty}(\frac{\partial H_0}{\partial t} +
[H_0,V])dt \ = \alpha
\sin(\psi) , \ \alpha = \frac{\pi
\varepsilon |1 -
\Lambda|\Omega^2}{\Lambda \sinh(\pi\Omega/2\sqrt{\Lambda-1})}, 
\end{equation} 
where $[...]$ denotes the Poisson bracket.

The period of the unperturbed motion is 
$T = 4\kappa K(\kappa)/(C\Lambda)$, where \cite{Shenoy2}:
\begin{equation}
\kappa^2 = \frac{1}{2}[1 + \frac{H_0 \Lambda - 1}{\sqrt{\Lambda^2 + 1 - 
2H_0 \Lambda}}] \quad , 
\end{equation}
\begin{equation}
 C^2 = \frac{2}{\Lambda^2}[(H_0 \Lambda - 1) +
\sqrt{\Lambda^2 + 1 - 2H_0 \Lambda}] \quad , 
\end{equation}
where $K(\kappa)$ is the complete elliptic integral of the first kind.
Near the separatrix, when $H_0 \rightarrow 1,$ we have  $ \kappa^2 \rightarrow
1$. Introducing the parameter $\delta = 1- H_0$, $H_0 <1$, $\delta << 1$,
and taking into account that $K(\kappa) \rightarrow
\ln(4/\sqrt{1-\kappa^2})$, when $\kappa^2 \rightarrow 1$, 
we obtain  the estimate for the period of oscillations near the separatrix
\begin{equation}\label{per}
T \approx \frac{2}{\sqrt{\Lambda-1}}\ln{\left(
\frac{4\sqrt{2(\Lambda-1)}}{\sqrt{\Lambda \delta}} \right) }.
\end{equation}

Using the expressions (\ref{MA}) and (\ref{per}) we find the whisker-type map
for our problem
\begin{eqnarray}\label{map}
\delta_{n+1} = \delta_n + \alpha \sin(\psi_n),\nonumber \\
\psi_{n+1} = \psi_n + \gamma
\ln\left( \frac{\mu}{\sqrt{\delta_{n+1}}}\right) . 
\end{eqnarray}
where $\gamma = 2\Omega/\sqrt{\Lambda-1}$ and $ \mu =
4\sqrt{2(\Lambda-1)}/\sqrt{\Lambda}$.

This map can be simplified, using the linearization around the fixed
points \cite{Reichl}. The fixed points are defined by 
\begin{equation}
2\pi l = \gamma \ln\left( \frac{\mu}{\sqrt{\delta^{(l)}_f}}\right), \ l=
1,2.. .
\end{equation}

Let us introduce the dimensionless energy $I_n$, defined by
\begin{equation}
I_n = -\frac{\gamma}{\delta^{(l)}_f} (\delta_n - \delta_f^{(l)} ).
\end{equation}

Substituting  this expression into Eq.(\ref{map}), and redefining $I_{n}
\rightarrow  I_{n}/2$,  we obtain the standard 
map
\begin{eqnarray}
I_{n+1} = I_n - K \sin(\psi_n ) \quad ,\nonumber \\
\psi_{n+1}= \psi_n + I_{n+1}\quad ,
\end{eqnarray}
where the parameter $K$ is
\begin{equation}
K = \frac{\alpha \gamma}{2\delta^{(l)}_f }.
\end{equation}
As is well known, chaos appears when $K \ge 1$ (exact value is $K^* =
0.9716$). Then we have the
estimate for the critical amplitude of modulation
\begin{equation}
\Lambda_1 \ge \frac{\delta^{(l)}_f \Lambda^3
\sinh(\pi\Omega/2\sqrt{\Lambda-1})}{\pi\sqrt{\Lambda-1}\Omega^3}.
\end{equation}
 
From the above, we may calculate the diffusion coefficient for the motion in the 
stochastic layer
\begin{equation}
D = \frac{<(\Delta E)^2 >}{t}.
\end{equation}
It results that we have $D = K^2/2$. For frequencies $\Omega \le
2\sqrt{\Lambda-1}/\pi$ we have the estimate
\begin{equation}
D \approx \frac{2^{10} \pi^2 \varepsilon^2 (\Lambda-1)^3
\Omega^4}{\Lambda^2}\exp{(-\frac{4\pi l\sqrt{\Lambda-1}}{\Omega})} \quad .
\end{equation}
For $\varepsilon = 0.1$, $\Omega = 2.5$, $\Lambda = 9$, $l = 1$, the time of 
exit from the stochastic layer is $10$ i.e. $\sim 4$ periods of the oscillations 
of the atomic scattering length.

\section{Rapidly varying scattering
length}

In the case when the frequency of oscillations of the atomic scattering
length is much larger  than the tunneling frequency $\omega_L = 2K$. i.e.
$\Lambda = \Lambda(t/\epsilon)$, where $\epsilon = \omega_L/\Omega << 1$,
it is
useful to derive an averaged set of equations. Using a multiscale approach 
\cite{Kevorkyan}  and expanding $z = \bar{z} + \epsilon z_1 +... , \Phi
=
\bar{\Phi} + \epsilon \Phi_1 ....$ we can
derive the system of equations for the slowly varying functions $\bar{z},
\bar{\Phi}$. The limits of validity is $\bar{z}^2 << 1$.

The averaged system turns out to be:
\begin{eqnarray}\label{av}
\bar{z}_t = -\sqrt{1 - \bar{z}^2}\sin(\bar{\Phi})(1 - \delta\bar{z}^2 ),\\
\bar{\Phi}_{t} = \nu <\Lambda >\bar{z} + \frac{\bar{z} \cos(\bar{\Phi}
)}{\sqrt{1-\bar{z}^2}}(1 + 2\delta - 3\delta \bar{z}^2 ).
\end{eqnarray}
where $\delta$ is proporttional to $\epsilon^2$, and for the harmonic
modulation is $\Lambda_1^2/(4\Omega^2 )$. The corresponding Hamiltonian is
\begin{equation}
H = \frac{\nu <\Lambda>}{2}\bar{z}^2 - \sqrt{1 -
\bar{z}^2}\cos(\bar{\Phi})(1 - \delta \bar{z}^2 )\quad .
\end{equation}

With respect to the unperturbed case, a new stable point can occur. A  
bifurcation is observed when 
$\delta$ is increased. The unstable point (in the undisturbed case) at $z= 0, 
\Phi = \pi$ may become stable under
rapid perturbations. This phenomenon is the analog of the Kapitza stabilization 
of the unstable fixed point of the pendulum by rapidly oscillating the 
suspension point. By passing though the bifurcation, a change in the topology of 
the phase portrait happens. The value of $\delta$  for this to occur
depends on 
$<\Lambda>$.  To have an estimate of the critical value of $\delta$, consider 
$z^2 << 1$ and $\Phi = \pi + \psi, \psi << \pi$. The equation for $z$ reads:
\begin{equation}
z_{tt} - (<\Lambda > - 1 - 2\delta )z = 0 \quad .
\end{equation}
Thus, when $\delta > \delta_c = (<\Lambda > -  1 )/2$ we have oscillatory
motion 
around $(0,\pi)$ in the phase plane. When $\delta < \delta_c$  we have an 
hyperbolic point, and $(0,\pi)$ is no longer stable. For example, with $\Lambda 
= 2$, the critical $\delta$ is $ 0.50$. For $<\Lambda> =1 $, the point is
always 
stable, as $\delta_c = 0$. To support these results, we have numerically 
integrated the original systems given by Eq.(\ref{tun}) with an explicit harmonic 
fast-time variation for $\Lambda(t)$. The above described features immediatly 
show up.  

Note also that we can cross the sign of $a_s$ from positive to negative. 
Experiments with very a large monotonic change of sign show that the BEC shrinks 
very fast. It would be interesting to check if the condensate
with attractive interaction of atoms  stable under the 
rapid variation of scattering length. This problem requires, however, separate 
investigations.  In any case in quasi $1D$ geometries collapse is
suppressed and this analysis is relevant.
 
In order to better understand the physical consequences of the averaging, we 
note that for small $\bar{z}^2 << 1$ we can rewrite the Hamiltonian in the form
$$H = \frac{\nu\Lambda_r z_r^2}{2} - \sqrt{1 - z_r^2}\cos{(\Phi)},$$
where $z_r = (1+\delta)\bar{z}$, $\Lambda_r = (1-2\delta )<\Lambda >$.
Thus we can conclude that the result of the averaging consists, for a
fixed initial phase difference between condensates, in the 
lowering of the critical relative population for self-trapping
regime to occured and in the increase of the critical value of the nonlinear
parameter $\Lambda$. We come to the picture of a more rigid pendulum in 
comparison with the constant case. The threshold for the
switching from the macroscopic quantum tunneling regime to the
self-trapping regime is shifted to lower initial values of the
atomic imbalance.

\section{Conclusion}

In this paper we have studied the new effects occuring due to
time-periodic variations  of the atomic scattering length. The 
resonances in oscillations of the atomic imbalance 
for the different regimes have been analyzed. 

The interaction of resonances occuring under oscillating scattering
length can lead to  complicated behavior of the atomic oscillations
and gives rise to the phenomenon of {\it chaotic macroscopic quantum
tunneling}.
In this work we prove the existence of such regime, using the Melnikov
function approach and calculating the Melnikov distance,
characterizing the width of the stochastic layer near the separatrix of the
unperturbed system. We derive a whisker type map for the problem and
obtain the estimate for the critical value of amplitude $\Lambda_1$
leading to the chaotic motion. The diffusion coefficient for the motion in
the stochastic layer is calculated and shown that the time to cross the
stochastic layer is the order of few periods of modulations.  
We also
consider the evolution of the system
under rapidly varying oscillations of the scattering length and derive the
averaged system for the slow-time variations of the  relative atomic population
$\bar{z}(t)$ and  the phase $\bar{\Phi}(t)$. The analysis of the fixed
points shows that the stabilization of the system under rapid
perturbation, when the unpertubed is unstable, is possible in the $\pi$
-phase mode.

\section{Acknowledgments}

F.Kh.A. is grateful to FAPESP (Brazil) and  INTAS (Grant 96-339) for
partial financial support,
and to Instituto de F\'{i}sica Te\'orica -- UNESP, S\~ao Paulo, for hospitality. 
R.A.K. thanks FAPESP and CNPq (Brazil) for partial financial support.

\newpage
\begin{figure}[t!]\label{Fig.1}
\centering
\hspace*{-5.5mm}
\leavevmode\epsfysize=8cm \epsfbox{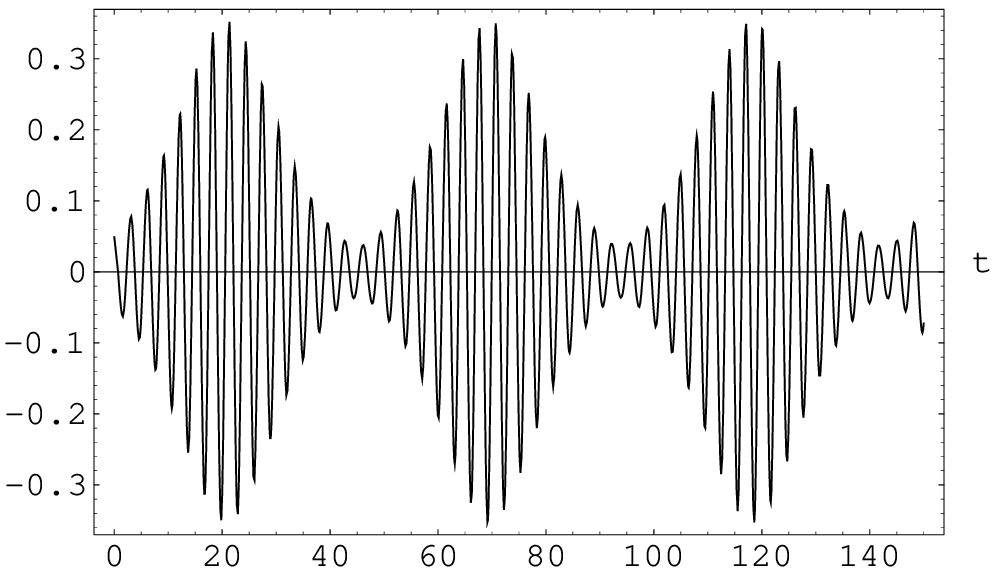}\\[3mm] 
\caption{The parametric resonance in zero-phase mode. The parameters
are:$\Lambda_0 = 3$, $\Lambda_1 = 0.12$, $\Delta = 0.1 $.}
\end{figure}
\newpage
\begin{figure}\label{Fig.2}
\centering
\hspace*{-5.5mm}
\leavevmode\epsfysize=8cm \epsfbox{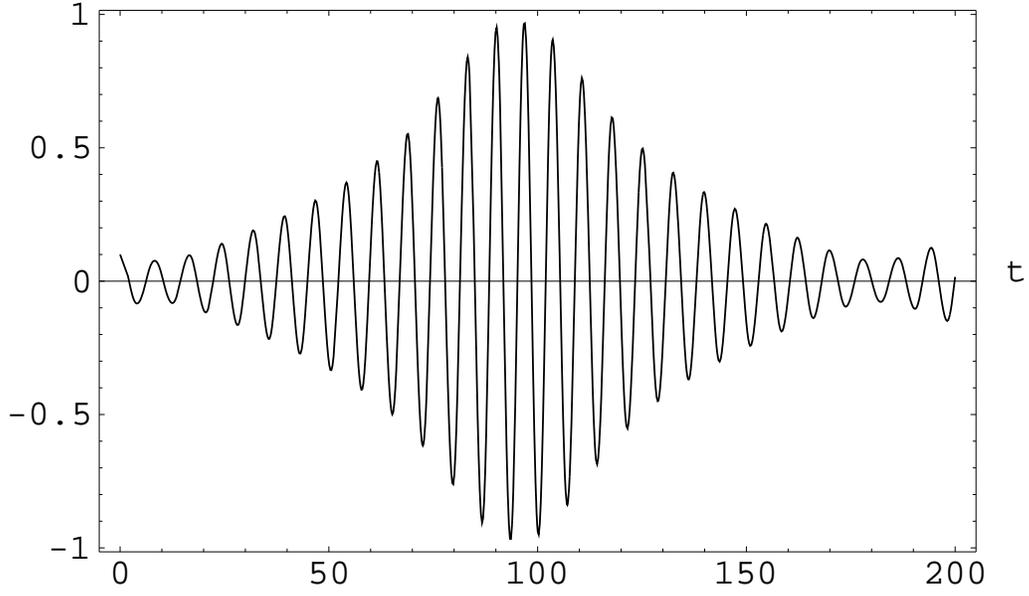}\\[3mm] 
\caption{The parametric resonance for $\pi$-phase mode: The parameters
are:$\Lambda_0 = 0.36, \Lambda_1 = 0.155, \Delta = 0.1, \Omega = 1.7,
\Phi(0) = \pi.$} 
\end{figure}
\newpage

\begin{figure}\label{Fig.3}
\centering
\hspace*{-5.5mm}
\leavevmode\epsfysize=8cm \epsfbox{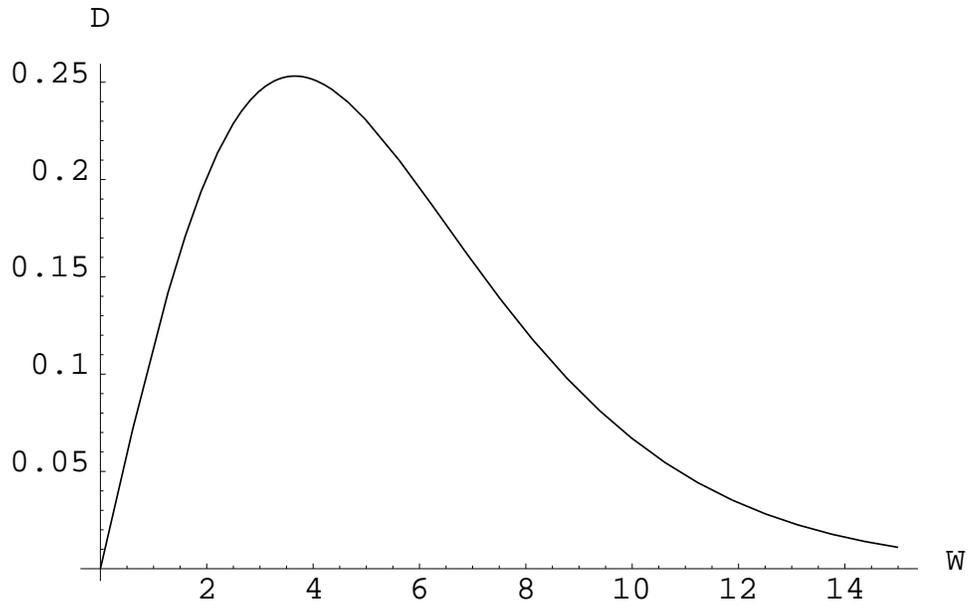}\\[3mm]
\caption{Melnikov $D(\Omega)$ distance versus frequency for $\epsilon =
0.1, \Lambda = 2, \phi(0) = 0$}
\end{figure}
\newpage
\begin{figure}\label{Fig.4}
\centering
\hspace*{-5.5mm}
\leavevmode\epsfysize=8cm \epsfbox{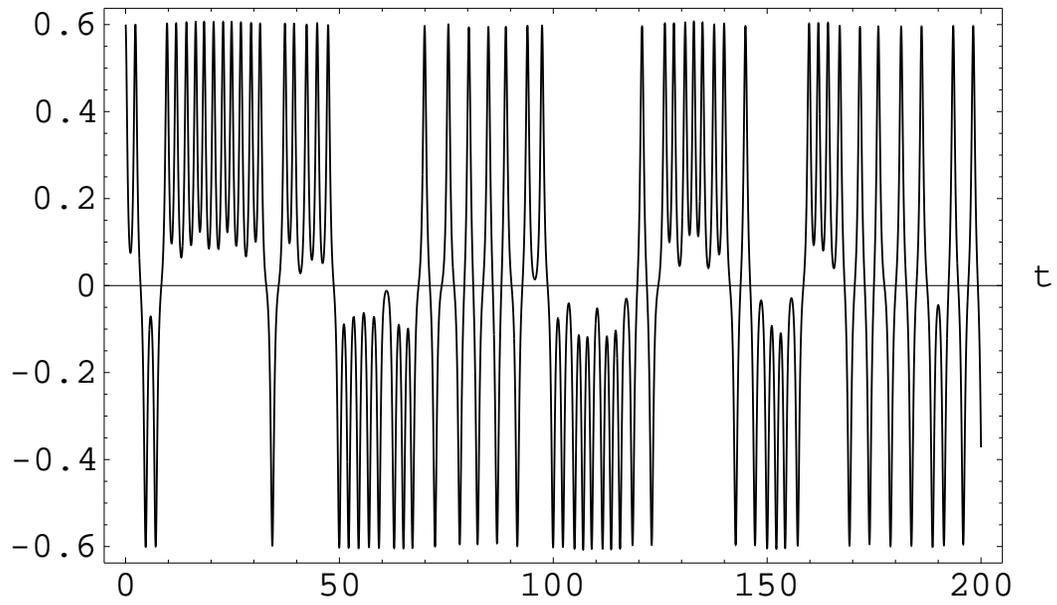}\\[3mm]
\caption{The chaotic oscillations of the relative atomic population
$z(t)$ for $ \Omega = 3.89, \Lambda = 2, \Lambda_0 = 0.2, \Phi(0) = 0, z(0) = 
0.6$}
\end{figure}
\newpage

%


\end{document}